\documentclass[a4paper]{article}
\pdfoutput=1
\usepackage{amsmath}
\usepackage{amssymb}
\usepackage{layout}
\usepackage{ifthen}
\usepackage{url}
\usepackage[pdftex,colorlinks=true,linkcolor=blue]{hyperref}
\usepackage[retainorgcmds]{IEEEtrantools}
\author{Roger Sewell}
\usepackage{graphicx}
\usepackage{xr}
\usepackage{float}

\title{A deterministic electoral system satisfying Arrow’s four
  conditions in an easily approached limit\footnote{First version
  submitted to arXiv; RFS version 1.9.1.1
}}
\author{Roger Sewell\\
\href{mailto:roger.sewell@cantab.net}{\scriptsize{roger.sewell@cantab.net}}}

\begin{document}
\maketitle

\begin{center}
\textbf{Abstract}
\end{center}

In 1950 Arrow famously showed that there is no social welfare function
satisfying four basic conditions. In 1976, on the other hand, Gibbard
and Sonnenschein showed that there does exist a unique probabilistic
social welfare method that satisfies a different set of strictly
stronger conditions. In this paper we discuss a deterministic
electoral method satisfying those same stronger conditions in an
appropriate sense; it is not a counterexample to either of these
theorems. We attach a simple reference implementation written in C
with executables for Linux and Windows.

\tableofcontents

\section{Introduction}

For many years Arrow's theorem\cite{Arrow}, which shows the
non-existence of electoral systems (precisely defined) satisfying four
basic conditions, has been taken by many\cite{McLean} as showing that
there are no electoral systems satisfying basic desirable properties,
and therefore as justification for the widespread deployment of
various systems with a variety of undesirable properties. Even after
Gibbard and Sonnenschein\cite{Gibbard1976} showed this is not the
case, many persisted with this view, dislike of the unique electoral
system (differently precisely defined) found by these authors and lack
of popular knowledge of their paper both playing their part.

We would like to point out that by modifying the definition of
``electoral system'' differently, but in a way still recognisable to
the general public as an electoral system, we can produce a system
satisfying not only Arrow's basic four conditions but a set of much
stronger conditions, at least in a rapidly approached limit and under
the assumption that at least two voters don't want to cede their
voting rights to the other voters. We would recommend this system as
fair and representative for electing either committees or
representatives in a parliament. 

We do not address the question of how to conduct business in a
parliament; the avoidance of ``tyranny by the majority'' in this different
setting is as far as we know an unsolved problem.

We first remind our readers of the details of the two crucial
background theorems. In section \ref{VDRD} we then define our proposed
system. In section \ref{properties} we discuss the conditions it
satisfies, before general discussion and conclusions in section
\ref{discussion}. In section \ref{implementation} we point to an
accompanying reference implementation.

\section{Background}
\label{background}

\subsection{Setting}

We consider the situation where each voter in an electorate expresses
a strong total ordering\footnote{An ordering $\leq$ on a set is
\textbf{total} if for all $x$ and $y$, either $x\leq y$ or $y\leq 
x$;
it is \textbf{strong} if ($x\leq y$ and $y\leq x$) implies $x=y$.}
on a subset of a set of candidates, all other candidates being
considered equal and less preferred. From a set of such votes we want
a system that outputs a strong total ordering on a subset of the
candidates of predetermined size $N_\text{places}$, all other
candidates being less preferred. We assume throughout that there are
more than two candidates.

\subsection{Arrow's Theorem}

Considering a slightly more restricted setting in which each voter
expresses a strong total ordering on all the candidates, Kenneth Arrow
showed \cite{Arrow} that no function exists that takes a set of such
votes as sole input, outputs a strong total ordering on the
candidates, and satisfies four very basic conditions:

\begin{description}

\item[Universal Domain]: The domain of the function includes the
  entire set of possible votes.

\item[No Dictator]: There is no voter whose vote is the output of the
  function irrespective of how the other voters vote.

\item[Weak Pareto]: If every voter prefers candidate $C_1$ to
  candidate $C_2$, then so does the output of the function.

\item[Irrelevant Alternatives]: Given two sets $S_1$ and $S_2$ of
  votes in which the voters ranking $C_1$ above $C_2$ are the same in
  both sets, the ranking of $C_1$ and $C_2$ in the outputs of the
  function on $S_1$ and $S_2$ are the same.

\end{description}

We will refer to these four conditions as UD, ND, WP, and IA.

This famous result is often taken as meaning that there are no decent
electoral systems\cite{McLean}. This, however, is
mistaken.

\subsection{The Gibbard-Sonnenschein Theorem}
\label{GibbardSonnenschein}

Still considering each voter expressing a strong total ordering on the
candidates, in 1976 Allan Gibbard and Hugo Sonnenschein showed
\cite{Gibbard1976} that there does exist a unique probabilistic
electoral system, known as Random Dictator (RD), satisfying the
following conditions:

\begin{description}

\item[Universal Domain]: The domain of the function includes the
  entire set of possible votes, and the output of the function is a
  probability distribution over the candidates (not over orderings of
  the candidates).

\item[Strategy Proof]: Whatever preferences over the candidates the
  voters may have, and however the other voters may have voted, the
  best option for a particular voter is to vote for his true
  preference order over the candidates.

\item[Symmetry on Candidates]: Every candidate is treated the same; in
  other words for any permutation of the candidates, if the voters'
  votes are unchanged, then so is the result (i.e. the probability
  distribution) given by the election.

\item[Symmetry on Voters]: Every voter is treated the same; in other
  words for any permutation of the voters, if each voter's vote is
  unchanged, then so is the result (i.e. the probability distribution)
  given by the election.

\item[Pareto Optimality]: If every voter prefers candidate $C_1$ to
  candidate $C_2$, then the probability of $C_2$ being elected (top)
  is zero.

\end{description}

Notice that any system that satisfies these conditions also satisfies
the four conditions of Arrow's theorem -- but the two differ in that
Arrow considers functions that output a total ordering on the
candidates, but Gibbard and Sonnenschein consider functions that
output a probability distribution over the candidates -- from which
the idea is that one then elects a single random sample from that
distribution.

We will refer to these conditions respectively as UD, SP, SC, SV, and
PO. 

\subsection{Random Dictator}

\subsubsection{The original}

The unique system identified by Gibbard and Sonnenschein operates as
follows:

\begin{enumerate}

\item Each voter votes for their favourite candidate (or gives a
  strong total ordering on the candidates).

\item One of the voters is chosen uniform-randomly from the
  electorate.

\item That voter's top preference is the result of the election.

\end{enumerate}

\subsubsection{An extension}
\label{RDextension}

This result can be extended to output a set of elected candidates of
predetermined size (so long as their are sufficient candidates, and so
long as sufficiently many of them are not boycotted by the entire
electorate). In such an extended system we proceed as follows:

\begin{enumerate}

\item Each voter votes a strong total ordering on a subset of the
  candidates, all others being considered equally unpreferred.

\item \label{execute} Execute Random Dictator (RD) to establish the
  first (or next) elected candidate as the favourite candidate of a
  uniform-randomly chosen voter, if his vote is non-empty.

\item Delete that candidate from the votes of all the voters.

\item Return to step \ref{execute} to elect another candidate unless
  we have already filled the desired total number of positions.

\end{enumerate}

Such an election will also fulfill UD, SP, SC, SV, and PO, regarding
each condition as applying stepwise to the election of each successive
candidate. 

\subsection{Voters' views on Random Dictator}
\label{dislike}

It is the author's personal experience that, despite its unique
attributes, RD is strongly disliked by most voters as a potential
electoral system. There seem to be two reasons for this:

\begin{enumerate}

\item \label{dislikerandom} They do not like the fact that randomness
  is involved, and in particular their inability to determine whether
  the random voter chosen was truly chosen at uniform-random from the
  set of voters.

\item \label{dislikeminorities} They do not like the fact that
  minorities have a chance of being represented; indeed the fact that
  it is a majority who choose whether to move away from e.g. the
  United Kingdom's ``First Past the Post'' electoral system which
  follows the majority's opinion on any \textbf{only-2-options} system
  is probably a major reason for the inability of the UK's electorate
  to decide to switch to a more representative system.

\end{enumerate}

It is the author's view that reason \ref{dislikerandom} has some
justification; the possibility of a random choice having been made,
but discarded by somebody who dislikes it and replaced by another, is
worrying. However reason \ref{dislikeminorities} does not seem
reasonable: everybody should have a fair chance of being represented.

\section{Voter-determined Random Dictator (VDRD)}
\label{VDRD}

We now turn to describing an electoral system that takes in voters'
strong total orderings on subsets of the voters (all others as before
being considered equally less preferred) and outputs a set of a
predetermined number of elected candidates (assuming the availability
of sufficient candidates not boycotted by every voter), and does so in
a deterministic manner.

\subsection{The essential idea}

The essential idea here is very simple: we implement RD, but instead
of a real random number (such as that chosen in the UK National
Lottery, or one chosen by observing radioactive decay), we use a
pseudo-random-number generator seeded by additional information
provided by each voter. We do this in such a way that so long as there
is at least one other voter whose seed contribution is equally likely
to take any of the permissible values, a voter can have no information
by which to decide what seed contribution would be to his advantage,
and hence we maintain strategy-proofness. Moreover there is a strong
disincentive to disclosing to other voters what seed contribution one
has voted: to do so would potentially hand power over the election to
the recipient of that information. However we stress that so long as
there are at least two voters who do not disclose their seed
contributions, the integrity of the election is maintained.

Because the input to the function executing the election is not just the
votes, but also seed information chosen by the voter, this approach is
not a counterexample to either Arrow's or Gibbard and Sonnenschein's
theorems.

\subsection{Choosing parameters}

The election organiser (who should be a trustworthy individual and
neither a candidate nor a voter) first chooses a ``large enough''
positive integer $K$. In practice we believe that $K=6$ is large
enough for many elections, as we will see, but some might wish to
choose e.g. $K=9$.

Define $M = 10^K$.

A pseudo-random-number generator with a long period is chosen; it
should preferably have a period much longer than the number of
possible orderings of the candidates, but certainly very much longer
than $C$, the number of candidates. It should be seedable by integers
in the range 0 to $M-1$.

\subsection{Definition}

A \textbf{seed contribution} is a $K$-digit decimal number, i.e. a
non-negative integer in the range 0 to $M-1$ inclusive.

\subsection{Producing the list of candidates}

In order to ensure that order of presentation of the candidates' names
does not influence voters, we start by ensuring that the list on the
ballot paper is in pseudo-random order.

To that end each candidate chooses a seed contribution and sends it to
the organiser along with his name. The candidate should not disclose
his seed contribution to others.

The organiser first orders the candidates in ascending order of their
seed contributions; two candidates with equal seed contribution are
ordered by alphabetical order of their names. This step ensures that,
when later a candidate is picked by a pseudo-random number from the
generator, the candidate so chosen is well-defined.

The organiser then adds the seed contributions modulo $M$ and seeds
the pseudo-random-number generator with the result. He then draws a
pseudo-random permutation of the candidates from the generator,
numbering them 1 to $C$, and prints the information for the voters
with the candidates numbered in this order.

\subsection{Voting}

Each voter provides both a seed contribution and an ordering of a
subset of the candidates in descending order of preference. Any
candidate not mentioned is assumed to be equally disliked by the voter
compared with all those mentioned.

\subsection{Determining the result of the election}
\label{determining}

The votes are sorted into numerical order of seed contribution, any
equal contributions being sorted by lexicographic ordering of the
numbers of the candidates voted for (the latter step being unlikely to
be necessary so long as $M$ is substantially bigger than $V^2$, where
$V$ is the number of voters). The votes (and corresponding voters) are
then numbered 0 to $V$ in that order.

The seed contributions of each voter are then summed modulo $M$ and
the result used to seed the pseudo-random-number generator.

We then execute the following steps:

\begin{enumerate}

\item Choose a voter using the pseudo-random-number generator.
\label{choosevoter}

\item Consider that voter's top preference candidate elected.

\item Delete that candidate from the votes of all voters, leaving the
  ordering of the remaining candidates listed by that voter unchanged.

\item Unless we have now elected enough candidates to fill all
  available places, return to step \ref{choosevoter}.

\end{enumerate}

\section{What conditions does VDRD satifsy ?}
\label{properties}

\subsection{Initial remarks}

It is clear that if we were to replace the pseudo-random-number
generator with a true random number generator then VDRD would be an
implementation of the extension of RD of section \ref{RDextension},
and would satisfy UD, SP, SC, SV, and PO, and hence also ND, WP, and
IA.

However, it is also clear that anybody armed with the relevant
pseudo-random-number generator and the votes can reproduce the result
of VDRD and hence check that the election has been validly executed --
unlike the situation with RD, where the random numbers chosen cannot
be reproduced or checked.

Further, it is trivial to check that VDRD itself satisfies UD and PO.

\subsection{Strategy-proof (SP)}

Since the seed contributions are summed modulo $M$, there are $M$
possible seeds that could be given to the pseudo-random-number
generator. 

Let us first suppose that there is at least one voter who does not
disclose his seed contribution to anybody, and who is equally likely
to have chosen any one of the $M$ possible seed contributions he could
make.

Then for any other voter and any seed contribution he may make, it is
equally likely that any of the $M$ possible random seeds may
result. Accordingly there is no strategy involving choosing a seed
contribution that will make any difference to the probabilities of the
various possible outcomes.

Alternatively, suppose that every other voter makes their seed
contribution known, and that without loss of generality they all
choose zero. Then the remaining voter can indeed have strategic
choices in his choice of seed contribution, and by trying out various
possible choices of seed in the publicly available
pseudo-random-number generator he can ensure that he himself is the
voter chosen in step \ref{choosevoter} of section
\ref{determining}. Thus for all the rest of the voters to reveal their
seed contributions is tantamount to giving dictatorial power to the
remaining voter.

So VDRD satisfies SP on the assumption that there are at least two
voters who don't wish to give up their voting power. (Indeed without
this assumption there is no need to hold an election at all.)

\subsection{Symmetry on candidates (SC)}
\label{VDRDSC}

It is clear that so long as all orderings of the candidates on the
ballot are equally likely, all candidates are treated
identically. (Indeed even if these orderings are not equally likely,
candidates are treated identically so long as no two voters use the
same seed contributions.) We therefore only need to show that each of
the possible orderings of the candidates on the ballot paper have
equal probability.

However, it is also clear that this cannot quite be true of VDRD: we
consider a specific scenario to illustrate.

Suppose $K=6$ and therefore $M=1000000$. Suppose there are 7
candidates. Then there are $7!=5040$ possible orderings of the
candidates, but one million possible seeds resulting from the various
seed contributions -- and one million is not divisible by 5040, so the
probability of getting each ordering cannot be equal.

But it is nonetheless worth asking how far from equality it is, and in
order to do that we need a way of measuring the difference between two
probability distributions, one the desired distribution, and the other
an approximation to it. It turns out that the natural way of doing
this is to use the Kullback-Leibler (KL) divergence \cite{KLdiv},
defined for a wanted distribution $P$ on a discrete variable $x$ and
an approximation $Q$ to $P$ to be the expectation of
$\log\left(\frac{P(x)}{Q(x)}\right)$ under $P$, or in this
case $$\sum_{x:P(x)>0}{P(x)\log\left(\frac{P(x)}{Q(x)}\right)}.$$ In general
this is the logarithm of the geometric average factor by which the
approximation is deficient; for small values it is roughly the average
fraction by which the approximation is deficient. Using natural
logarithms to base $e$ the KL divergence is measured in nats.

In this specific case, using the pseudo-random-number generator from
the implementation of section \ref{implementation}, it turns out that
the KL divergence of the resulting distribution on the 5040 orderings
from the uniform distribution on them is about 0.00251 nats; that is
to say that the average fractional error in these probabilities is
about 0.25\%.

We may also be interested in the resulting probability distributions
on the candidate being ordered in first place on the ballot paper (or
2nd, 3rd, etc). In this specific case these KL divergences are all
less than $3\times 10^{-6}$ nats, so the probabilities are in error by
on average less than 0.0003\%.

It is intuitive, but probably unprovable in practice for any actual
set of pseudo-random-number generators, that as both $K$ and the
period of the generator go to infinity with the number of candidates
fixed, so the KL divergences on both orderings and places will tend to
zero. From the above discussion we see that for 7 candidates $K=6$ is
adequate to achieve SC for practical purposes. For a theoretically
perfect pseudo-random-number generator with infinite period the strong
law of large numbers and the finiteness of the number of orderings
will ensure that the KL divergences go to zero as $K\to\infty$.

We should also mention a difficulty that arises for larger numbers of
candidates. If $C=10$ but still $K=6$ there are $10! = 3628800$
possible orderings of the candidates, but only one million possible
seeds. It is then inevitable that at least two million possible
orderings have probability zero of occurring, and then since on such
an ordering $Q$ is deficient by an infinite factor compared with the
wanted uniform distribution, the KL divergence will be $\infty$. We
would, however, still like to get a measure of how bad the
approximation is to to the desired distribution. For this we instead
measure the KL divergence between two modified distributions: in both
the wanted and the approximation cases we substitute a mixture of $M$
parts of the distribution in question and $|X|!$ parts of the uniform
distribution, where $X$ is the set of orderings or the set of
candidates as appropriate. The difference between the modified and
unmodified KL divergences goes to zero as $K\to \infty$.

In the case of $C=10$ and $K=6$ the KL divergence on orderings of the
candidates will be $\infty$ and the modified KL divergence turns out
to be 0.063 nats on orderings, while even the unmodified KL divergence
is less than $10^{-5}$ nats on all places.

Thus given a theoretically perfect pseudo-random-number generator, SC
holds in the limit as $K\to\infty$, and in practice $K=6$ is adequate
for up to 7 candidates. Even with 10 candidates and $K=5$ the KL
divergence for the \textit{place} distributions is below $1.2\times
10^{-5}$ nats.

\subsection{Symmetry on voters (SV)}

Essentially the same considerations apply as discussed in section
\ref{VDRDSC}. With a perfect pseudo-random-number generator the KL
divergence of the distribution on output orderings of the election
from the theoretical RD distribution goes to zero as $K\to\infty$, and
RD satisfies SV. In practice with 5 candidates, 100 voters, and random
votes, K = 6 gives KL divergence of less than $10^{-4}$ nats on output
orderings of the election from the theoretical distribution, while
that for places is under $3 \times 10^{-6}$ nats.

\subsection{Summary: conditions satisfied by VDRD}

We make the following assumption:

\begin{description}

\item [No Yielding]: At least two candidates and at least two voters
  do not reveal any information about their seed contribution(s)
  before all seed contributions have been chosen and recorded.

\end{description}

Then in the limit as $K\to \infty$ with a perfect pseudo-random-number
generator VDRD satisfies UD, SP, SC, SV, and PO, hence also ND, WP,
and IA, all considered stepwise as each successive candidate is
elected. For a practical pseudo-random-number generator, $C=5$, and a
random set of 100 votes, already for $K=6$ the KL divergence of the
resulting distribution on the orderings output by the election from
the true RD distribuion is only $4.4 \times 10^{-5}$ nats, while the
corresponding figures for places are all under $3 \times 10^{-6}$
nats.

\section{Discussion}
\label{discussion}

There are, of course, many possible ways of incorporating seed
information into votes, any of which could give a variant of VDRD. The
method suggested here has the advantage of satisfying SP under the No
Yielding assumption -- and if the No Yielding assumption were false,
then no election would be needed. 

As for the need for $K$ to approach infinity for SC and SV to hold, we
note first that for any practical pseudo-random-number generator there
will be a finite seed space, and with that fixed the probability
distribution realised on output orderings will not in general exactly
match the RD distribution. We believe the approximation offered by
VDRD is entirely adequate for real-life elections, particularly as
setting $K=6$ is both sufficient not only to get a good approximation
to the places distributions, but also to the distribution on orderings
for $C\leq 6$ -- and it is entirely practical for a voter to choose a
six-digit number.

Our hope is that the deterministic nature of VDRD will overcome
dislike reason \ref{dislikerandom} of section \ref{dislike}. It is
obviously not our intention to remove the factors leading to reason
\ref{dislikeminorities}, as not to give minorities representation
would in our view totally defeat the purpose of the exercise.

In general choosing an electoral system on the basis of the desirable
properties that it satisfies seems to us an appropriate approach. In
this case RD is the unique probabilistic system satisfying the
conditions listed in section \ref{GibbardSonnenschein}, and VDRD
approximates it arbitrarily well as $K\to\infty$, and in practice well
enough for $K=6$. We believe that this list of conditions are all
eminently desirable. 

Two things remain:
\begin{enumerate}

\item Convince people that ``tyranny by the majority'' is a bad thing
  and that choosing an electoral system on the basis of the axioms it
  satisfies is a good idea; and

\item Work out a way of conducting business in an elected parliament
  or committee that satisfies a suitable set of desirable properties
  and that avoids ``tyranny by the majority''.

\end{enumerate}

\section{Reference implementation}
\label{implementation}

We provide with this arXiv paper a reference implementation that can
be run on Linux and on Windows7 to Windows10 (and maybe on
Windows11). Its intent is simply to provide a means of checking that
any more user-friendly implementation carries out the VDRD procedure
in exactly the same well-defined manner, in order that the result of
an election can be checked by anybody using any such implementation.

For arXiv-related technical reasons the software is provided as a
Linux file \texttt{rndict.v1.26.tar.xz} which can be downloaded from
the same page of arXiv on which the abstract is shown (see the panel
on the right). On Windows 10 and on Linux the file can be unpacked by
using the Terminal (Command Prompt), changing to the relevant
directory, and typing \texttt{tar -xJf rndict.v1.26.tar.xz}$\ $ --
\textbf{on Windows you can ignore any ``Cannot create symlink'' errors
  in the subdirectory \texttt{rndict/rndictC/Linux64}} -- then read
the \texttt{DO\_READ\_ME.pdf} file. If you are running an earlier
version of Windows, and do not have the \texttt{tar} command
available, the easiest way to unpack it is probably to ask a friend
who is running Windows 10 or Linux to unpack it for you, or to use
7-Zip if you already have it installed.

\bibliography{ms}

\begin{thebibliography}{1}

\bibitem{Arrow}
K.~J. Arrow, ``A difficulty in the concept of social welfare,'' {\em The
  Journal of Political Economy}, vol.~58, no.~4, pp.~328--346, 1950.

\bibitem{McLean}
I.~McLean, {\em Public choice: an introduction}.
\newblock Basil Blackwell, 1987.
\newblock See page 168.

\bibitem{Gibbard1976}
A.~Gibbard, ``Manipulation of schemes that mix voting with chance,'' {\em
  Econometrica}, vol.~45, no.~3, pp.~665--681, 1977.

\bibitem{KLdiv}
S.~Kullback and R.~A. Leibler, ``{On Information and Sufficiency},'' {\em The
  Annals of Mathematical Statistics}, vol.~22, no.~1, pp.~79 -- 86, 1951.

\end{thebibliography}
\bibliographystyle{ieeetr}

\end{document}